# Comparative Analysis of the Main Nasal Cavity and the Paranasal Sinuses in Chronic Rhinosinusitis: An Anatomic Study of Maximal Medical Therapy

Satyan B. Sreenath, MD[1], Julia S. Kimbell, PhD[1], Saikat Basu, PhD[1], Andrew J. Coniglio, MD[1], Tatayana E. Fontenot, MD[1], Brian D. Thorp, MD[1], Charles S. Ebert, MD, MPH[1], Brent A. Senior, MD[1], Adam M. Zanation, MD[1,2]

[1]Department of Otolaryngology—Head and Neck Surgery, University of North Carolina at Chapel Hill

[2]Department of Neurosurgery, University of North Carolina at Chapel Hill

**Corresponding Author:** Adam M. Zanation, MD, Department of Otolaryngology—Head & Neck Surgery**,** Department of Neurosurgery, University of North Carolina at Chapel Hill

Address: 170 Manning Drive**,** CB #7070, Physician's Office Building Room G-190, Chapel Hill, NC, 27599

Phone: 919-966-3342

Fax: 919-933-7941

Email: adam_zanation@med.unc.edu

**Financial Disclosure: None**

**Conflict of Interest: None**

## Abstract

### Objective

Minimal literature exists investigating changes in inflammation with respect to the main nasal cavity (MNC) and paranasal sinuses (PS) before and after maximal medical therapy (MMT) for chronic rhinosinusitis (CRS). We hypothesized that MMT produces a differential level of change in the volume of air space in the MNC and PS, and that resolution of mucosal disease associated with the osteomeatal complex (OMC) influences clinical response to MMT.

### Study Design

Retrospective study of 12 pre- and post-MMT sinus-CT scans from 6 subjects with CRS, of which three succeeded and three failed therapy.

### Methods

Mimics™ was used to create 3D-models of the MNC and PS, and then analysis of the models was performed.

### Results

Mean differences in the sinonasal volume were $7866.5 \pm 4339.9$ mm$^3$ and $17869.10 \pm 19472.70$ mm$^3$, amongst the failures and successes, respectively. There is wide variability in the contribution of PS and MNC airspace volume change to the overall change in the sinonasal volume. In two subjects, the direction of volume change in the MNC and PS diverged with respect to the overall change in volume. Line-of-sight analysis demonstrated that successful responders to MMT had more patent MNC with direct access to the OMC.

### Conclusions

There is a differential contribution to sinonasal, airspace volume change after MMT, when comparing the MNC and PS. Response to MMT may not be solely attributable to PS change and may include a function of MNC change. Line-of-sight models suggest that direct access to the OMC may impact response to MMT.

**Level of Evidence: 2b**

## Introduction

Before consideration of functional endoscopic sinus surgery (FESS), it is generally accepted that patients with chronic rhinosinusitis (CRS) undergo a trial of "maximal medical therapy" (MMT).[1-3] Although variation exists with regard to the definition and duration of MMT, the mainstay of treatment focuses on oral antibiotics, topical intranasal steroids, and sinonasal saline irrigation, with possible systemic oral steroid therapy.[1-3] In a survey, greater than half of otolaryngologists who responded used sinonasal saline irrigation, intra-nasal steroids, oral steroids, and oral antibiotics with widely varying treatment durations in greater than 50% of patients presenting for initial evaluation of CRS.[1] Though many studies have confirmed the efficacy of MMT in the management of CRS, there is no agreed upon consensus regarding the most effective regiment.[4-7]

The goals in the treatment of CRS are reduction in inflammation of the sinonasal mucosa, improved mucociliary clearance of the paranasal sinuses, and enhanced sinonasal airflow through the main nasal cavity with associated improvement in clinical symptoms.[2,4,8-11] Currently, there are a number of studies that have investigated the effect of FESS on the sinonasal cavity and implications for nasal irrigation and drug delivery. According to Wofford et al., who studied the effect of FESS on topical maxillary sinus drug delivery, FESS with larger maxillary antrostomies allowed for improved topical drug delivery.[12] Additionally, Frank et al. confirmed the enhancement in airflow particularly in the maxillary sinus that was associated with improvement in patient quality of life as well based on symptom questionnaires.[13]

The majority of literature in the study of the sinonasal cavity of CRS patients has primarily focused on pre- and post-surgical evaluation of the main nasal cavity (MNC) and paranasal sinuses (PS). However, scant literature exists investigating changes in the MNC and PS before and after MMT for CRS, and whether such changes are differential. In addition, it is unclear whether certain disease patterns influence clinical success or failure of MMT. According to Lal et al., clinical predictors of failure of medical therapy for patients with CRS included higher patient scores of facial pressure and significant endoscopic mucosal inflammation.[4] However, given that computed-tomography (CT) analysis generally provides the most comprehensive and objective method of assessing sinonasal disease severity, assessment of sinus CT's can often provide important information regarding change after treatment intervention.[14]

Therefore, we sought to differentially compare and analyze changes in the MNC and PS using three-dimensional (3D) reconstructed models of sinus CT scans of patients obtained before and after MMT for CRS. In this preliminary study of six patients, we hypothesized that there is a differential level of change in the volume of air space and mucosal inflammation in the nasal cavity when compared to the paranasal sinuses, and that these volumes of change distinctively impact the overall change in air space after medical therapy. Secondly, we hypothesized that resolution of mucosal disease associated with the osteomeatal complex (OMC) influences clinical response to MMT and direct visualization of the OMC from the external nares may be associated with success of MMT.

**Materials and Methods**

After approval from the Institutional Review Board, sinus-CT scans from subjects with CRS who had been treated with MMT were obtained. Patients from the subject pool had been diagnosed with CRS based on clinical criteria and none had prior surgical therapy.[9] Of note, patients in the subject pool were previously enrolled in a prospective, randomized cohort study evaluating MMT, however this study is a retrospective analysis of the sinus CT scans that were obtained as part of the prior study protocol.[5] MMT for these patients was defined as an oral antibiotic course, intra-nasal steroid treatment with Flonase, and sinonasal saline irrigation. Prior to initiation of treatment and at the end of the treatment course, sinus CT scans were obtained. Additionally, rhinosinusitis disability index (RSDI) scores were obtained pre and post-therapy as a subjective measure of symptom improvement, and Lund-McKay (LM) scores were obtained from all sinus CT scans as part of an objective measure of change.

A total of 12 CT scans were chosen for this preliminary evaluation. Of these 12 scans, 6 sinus CT scans represented pre and post-therapy scans from patients who were deemed clinically successful responders to MMT. Conversely, the remaining 6 sinus CT scans represented pre- and post-therapy scans from patients who were deemed clinical failures to MMT and would require surgical intervention. Successful response or failure to MMT was determined based on clinical decision utilizing subjective and objective scoring measures including RSDI and LM scores, in addition to global patient symptoms.

Next, Mimics 18.1™ (Materialise, Inc., Plymouth, MI, USA) was used to create anatomically realistic three-dimensional (3D) reconstructions of the MNC and PS airways from the pre- and post-therapy CT scans. In each scan, the airspace was selected by setting a maximum threshold of -300 to -351 Hounsfield units based on visual inspection. The resulting pixel selection was then hand-edited to achieve anatomical accuracy and to separate the PS from the MNC (Fig. 1). Volumetric analysis was then performed on the nasal air spaces of the pre- and post-therapy sinus CT 3D models, as a negative correlate of airway obstruction, purulence, and swelling associated with mucosal inflammation. Of note, prior to the volumetric study, the effect of nasal cycling in the CT analysis was accounted for by comparing the pre and post-CT scans and ensuring for no unilateral changes that may confound analysis. Mimics™ was also used to coregister pre-and post-therapy sinus CT scans for direct comparison of overlaid 3D reconstruction contours to assess direct changes in MNC structures including the turbinates and access to the osteomeatal complex (OMC).

The 3D reconstructions were then exported from Mimics™ in stereolithography file format and imported into the computer-aided design and meshing software ICEM-CFD™ 15.1 (ANSYS, Inc., Canonsburg, PA, USA). Within this software, study of the OMC was performed via "line-of-sight" analysis. The "line-of-sight" analysis was performed by computational rotation of the sinus CT models to re-create how and if the OMC could be visualized via the external nares. The centroid (center-of-mass) of each of the bilateral nasal planes was first obtained. With the head in an upright position, each side of the nasal vestibule was cut by a horizontal plane (roughly parallel to the hard palate) at a positive vertical distance of 5 mm from the nostril centroid on that side (Fig 1). This distance simulated the viewpoint of a nasal speculum or the approximately position of the tip of a nasal spray bottle. The bilateral OMCs were marked with points

(Fig. 1) and visualization of these points from the cut vestibular areas was used to compute a line-of-sight grade (Fig. 1). An objective scoring system was devised on scale of 1-5. A score of 1 indicated full visualization of the OMC, where as a score of 5 indicated zero visualization of the site. Varying degrees of visualization were scored in between with the lower numbers correlating with better visualization. Each pre-treatment model was subject to a "line-of-sight" analysis from the left nare and the right nare. The scores were combined and compared between the medical successes and failures.

**Results**

The selected cohort included three clinical successes and three clinical failures of MMT for CRS. RSDI and LM scores worsened or remained the same after MMT in three subjects (Table 1). Conversely, RSDI and LM scores overall improved or remained the same after MMT in three subjects (Table 1). Importantly, all failure patients in the cohort required FESS to address CRS symptoms.

There was a mean increase in the sinonasal airspace volume in the clinical success group when combining the volume of the MNC and PS and a mean decrease in the clinical failures group (Tables 2, 3). Using 3D reconstructive techniques, this change was captured and used to analyze the MNC and PS separately (Fig, 2). In the three failures, there was an average decrease in the airspace volume of the MNC and PS of $4043.4 \pm 3092.9$ and $4118.7 \pm 6331.5$, respectively (Table 3). In the three successes, there was an average increase in the airspace volume of the MNC and PS of $3805.8 \pm 8046.0$ and $13701.5 \pm 17574.8$, respectively (Table 3). Of note, in each outcome group, two of the three subjects had consistent and expected directions of volume change of the MNC and

PS.  However, in each outcome group, there was one subject whose direction of volume of change in the MNC and PS diverged with respect to the overall change in the sinonasal airspace volume.  In the success group, the volumetric analysis of subject 6, whose total sinonasal volume increased, demonstrated an unexpected decrease in the MNC volume, and in the failure group, the volumetric analysis of subject 1, whose total sinonasal airspace volume decreased, demonstrated an unexpected increase in the PS volume (Table 2).   Of note, when individually examining each patient, there is wide variability in the contribution of PS and MNC airspace volume change to the overall change in the sinonasal airspace volume.  With regard to the nasal turbinates, changes in turbinate hypertrophy were noticed throughout, but did not correlate with clinical response to MMT.

From the overlay analysis, OMC patency was achieved or there was noticeable reduction in inflammation surrounding the complex itself in the success group (Fig. 3A). However, despite MMT, there was worsening of mucosal obstruction at the OMC or no improvement in mucosal inflammation surrounding the OMC in the three failures (Fig. 3B).

Visualization of the OMC using digitally created "line-of-sight" demonstrated that successful responders to MMT had more patent MNC with direct access to the OMC (Fig. 4). The three patients with medical success had line-of-sight scores of 2, 5, and 7 with an average score between the three of 4.67 (Table 4). As anticipated, those who failed medical management had higher line-of-sight scores of 6, 9, and 9 with an average score of almost double, 8.0 (Table 4).

**Discussion**

Although there is a general consensus that MMT should be attempted prior to consideration of FESS, a recent systematic review showed that there is no definitive guideline on what constitutes MMT and only 21% of included studies demonstrated explicit MMT criteria that are necessary to fail prior to offering FESS.[8] Additionally, though there is demonstrated benefit, the direct impact of MMT on the sinonasal cavity remains unclear.[11] In conjunction, few studies have addressed differential impact of medical therapy for CRS on the MNC when compared to PS, and whether certain disease patterns lead to more medically refractory disease.

3-D reconstructions of sinus CT scans from six CRS patients before and after MMT allowed for review of specific anatomic changes in the sinonasal cavity prior to MMT. The total sinonasal volume increased in all three clinical successful responders and decreased in all three clinical failures. These volumetric changes are consistent with predicted CT changes of the two cohorts, as generally enlargement of the airspace volume of the sinonasal cavities post-therapy is indicative of decreased mucosal inflammation and thus improvement in sinonasal symptoms as seen on RSDI scores. However, once the PS were separated from the MNC, there appeared to be an interesting divergence in findings in addition to differential impact of each the MNC and PS to the overall volume change.

First, amongst the success group, the direction of volume change of the PS and MNC was congruent with the general increase in overall sinonasal volume in two subjects. However, even amongst these two subjects, there is a notable difference in how much PS compared to the MNC volume change contributes to the overall sinonasal volume change, wherein one subject the PS volume change is the dominant factor compared the MNC volume change being the dominant factor for the other patient. When comparing the pre and post-therapy LM and RSDI scores for these patient, there appeared to be a greater improvement in the RSDI score for the patient whose PS volume change seemed to be primary driver of sinonasal volume increase compared to the other patient in MNC volume change was the primary driver.  Interestingly, in the third success patient with diverging changes, whose PS volume increased while the MNC volume decreased, the total sinonasal volume change was lower, however, the patient's symptom scores on the RSDI demonstrated significant improvement.  Although limited by sample size, this data suggests that there may be a differential impact of MMT on the PS and MNC.  Given incongruity amongst the patients' changes, it is difficult to state whether changes in the MNC vs. the PS have a more clinical impact on response to MMT.

      Similarly, in the failure group, the direction of volume change of the PS and MNC was congruent with the general decrease in overall sinonasal volume in two subjects. Again, even amongst these two subjects, there is a notable difference in how much PS compared to the MNC volume change contributes to the overall sinonasal volume change.  Interestingly amongst these two subjects, there was no difference in the pre- and post-therapy LM scores, showing that the sensitivity of the LM score to subtle sinonasal changes is lower.  The third patient of the failure cohort also interestingly demonstrated diverging changes in the volume of the MNC compared to the PS with respect to overall sinonasal volume change, wherein there was a small increase in PS volume compared to a larger decrease in MNC volume that was the primary driver. This data demonstrates differential impact of MMT on the MNC vs. the PS, but without significant antecedent

literature investigating the direct impact of MMT on the sinonasal cavity, it is difficult to assess how this impacts response to medical treatment.

The osteomeatal complex plays a pivotal role in CRS given its physiologic function as a common, sinus drainage pathway and association with increased disease burden.[15] Upon performing overlay analysis, amongst the clinical failures, there was either no change or clinical worsening of the mucosal inflammation in the MNC, particularly at the level of the OMC. However, amongst the successful responders, there was either no change or clinical improvement in the mucosal inflammation surround the OMC.

Lastly, as intranasal steroid spray is an essential component of MMT, the "line of sight" analysis provided information regarding the effect of medical management of CRS. Although the sample sizes of the cohorts are low, improved visualization of the OMC from the nares theoretically increases the effectiveness of MMT when considering the use of intranasal steroid sprays. Also, given that all of the three successful patient responders to MMT demonstrated improved "line of sight" to the OMC whereas failure subjects demonstrated poor "line of sight" to the OMC, based on study of the pre-operative CT scans, a direct route for drug application may predispose patients to increased responsiveness to MMT and less likely to require operative intervention.

Importantly, there are limitations to note. Given the small sample size in both cohorts, it is difficult to make population based inferences about responsiveness to MMT. Secondly, when separating the PS from the MNC, there is no standardized method of separating sinuses from the nasal cavity, thus there is some variability noted between 3D reconstructions. After denoting differential change in the MNC and PS volume in patients after MMT for CRS, the next step in elucidating the effect of MMT on the sinonasal cavity is to perform computational fluid dynamics (CFD) analysis on these patients' CD.[12,16] Through CFD, differential change on the nasal airflow within the MNC and PS can be studied to determine if MMT has a direct impact on nasal airflow and how clinical responders differ from clinical failures.

## **Conclusion**

This study demonstrated that MMT has a definitive impact on changes in the sinonasal cavity, including volumetric changes in the airspace of the MNC and PS and inflammation surrounding the OMC, a critical determinant of disease burden in CRS. Of note, there appears to be a differential contribution to the sinonasal, airspace volume change after MMT, when comparing the MNC and PS. Though volumetric analysis suggested that the MNC changes as well as sinus patency, this should be further studied for potential differential changes in a larger cohort. Overlay analysis showed that OMC patency improvement was consistent with good patient outcomes of MMT. Thus, success or failure of MMT may not be solely attributable to paranasal sinus change and may include a function of nasal cavity change. Lastly, line-of-sight analysis suggested a potential new predictor of MMT success in patients presenting with CRS, however, further investigation of anatomic changes after MMT is required in a larger study group.

## **Acknowledgements**

Research reported in this publication was supported by the National Heart, Lung and Blood Institute of the National Centers of Health under award number

**Figure and Table Legend**

**Figure 1: Line of Sight Methodology**

Line-of-sight models demonstrating methodology for calculating OMC visualization. 1A. Line-of-sight analysis of the left nostril shows an example of N(Y) or approximately 50% visualization of the osteomeatal complex (OMC). 1B. There is a fragment of OMC visible within the left nostril (arrow); this was scored as a (N). 1C: Full visualization of bilateral OMC, scored as Y bilaterally."

**Figure 2: 3D-Reconstructions of the Airspace of the Sinonasal Cavity**

3D-reconstructions of the air space of the nasal cavity and paranasal sinuses from pre- and post-therapy sinus CT-scans that were performed on patients who had undergone maximal medical therapy (MMT) for CRS. 1A. Reconstructed models from subject 1 that demonstrates decrease in sinonasal volume in patient deemed clinical failure. 1B. Reconstructed models from subject 4 that demonstrate an increase in sinonasal volume in patient deemed successful responder to MMT.

**Figure 3: Overlay Analysis of the Airspace of the Main Nasal Cavity**

Using overlay analysis of the airspace of the main nasal cavity, the airspace outlines of the pre- and post-therapy CT scans were superimposed on each other to demonstrate differential change in mucosal inflammation and the turbinates. 1A. Clinical worsening of mucosal inflammation in the main nasal cavity and OMC, where pre-therapy (red) outline demonstrates greater airspace volume compared to the post-therapy (blue) outline. 1B. Clinical improvement in mucosal inflammation in the main nasal cavity and towards the OMC, where post-therapy (blue) outline demonstrates greater airspace volume compared to the pre-therapy (red) outline.

**Figure 4: OMC Visualization via Analysis of 3D-Reconstructions**

Manipulation of 3D-reconstructions demonstrating view of the OMC from the external nares. Upon manipulation of the 3D-models, it was seen that patients with successful response to MMT had a direct line-of-sight and access to the OMC when compared to patients who had failed MMT. 4C. Direct view of the mucosal region overlying the OMC (dots) in subject 6, a successful, clinical responder to MMT. 4D. No direct visualization of the sinonasal region overlying the OMC in subject 1, a clinical failure after MMT.

**Table 1: RSDI and LM Scores of Subject Cohorts of Failures and Successful Responders Clinical Responders to MMT**

RSDI- Rhinosinusitis Disability Index. LM- Lund-Mackay

**Table 2: Total Volumes of the Main Nasal Cavity and the Paranasal Sinus Air Space Before and After MMT in the Clinical Success and Clinical Failure Groups**

**Table 3: Mean Differences Between the Pre- and Post-Therapy Sinonasal Airspace Volumes of the Clinical Success and Clinical Failure Groups**

**Table 4: Line of Sight Analysis**

LHS - Left Hand Side; RHS - Right Hand Side; Y - Full Visualization of the OMC; (Y) - Greater than 50% of OMC visualized; N(Y) - Half of OMC visualized; (N) - Less than 50% of OMC visualized; N - No visualization of the OMC

**Figure and Table Legend**

**Figure 1: Line of Sight Methodology**

Line-of-sight models demonstrating methodology for calculating OMC visualization. 1A. Line-of-sight analysis of the left nostril shows an example of N(Y) or approximately 50% visualization of the osteomeatal complex (OMC). 1B. There is a fragment of OMC visible within the left nostril (arrow); this was scored as a (N). 1C: Full visualization of bilateral OMC, scored as Y bilaterally."

**Figure 2: 3D-Reconstructions of the Airspace of the Sinonasal Cavity**

3D-reconstructions of the air space of the nasal cavity and paranasal sinuses from pre- and post-therapy sinus CT-scans that were performed on patients who had undergone maximal medical therapy (MMT) for CRS. 1A. Reconstructed models from subject 1 that demonstrates decrease in sinonasal volume in patient deemed clinical failure. 1B. Reconstructed models from subject 4 that demonstrate an increase in sinonasal volume in patient deemed successful responder to MMT.

**Figure 3: Overlay Analysis of the Airspace of the Main Nasal Cavity**

Using overlay analysis of the airspace of the main nasal cavity, the airspace outlines of the pre- and post-therapy CT scans were superimposed on each other to demonstrate differential change in mucosal inflammation and the turbinates. 1A. Clinical worsening of mucosal inflammation in the main nasal cavity and OMC, where pre-therapy (red) outline demonstrates greater airspace volume compared to the post-therapy (blue) outline. 1B. Clinical improvement in mucosal inflammation in the main nasal cavity and towards the OMC, where post-therapy (blue) outline demonstrates greater airspace volume compared to the pre-therapy (red) outline.

**Figure 4: OMC Visualization via Analysis of 3D-Reconstructions**

Manipulation of 3D-reconstructions demonstrating view of the OMC from the external nares. Upon manipulation of the 3D-models, it was seen that patients with successful response to MMT had a direct line-of-sight and access to the OMC when compared to patients who had failed MMT. 4C. Direct view of the mucosal region overlying the OMC (dots) in subject 6, a successful, clinical responder to MMT. 4D. No direct visualization of the sinonasal region overlying the OMC in subject 1, a clinical failure after MMT.

**Table 1: RSDI and LM Scores of Subject Cohorts of Failures and Successful Responders Clinical Responders to MMT**

RSDI- Rhinosinusitis Disability Index. LM- Lund-Mackay

**Table 2: Total Volumes of the Main Nasal Cavity and the Paranasal Sinus Air Space Before and After MMT in the Clinical Success and Clinical Failure Groups**

**Table 3: Mean Differences Between the Pre- and Post-Therapy Sinonasal Airspace Volumes of the Clinical Success and Clinical Failure Groups**

**Table 4: Line of Sight Analysis**

LHS - Left Hand Side; RHS - Right Hand Side; Y - Full Visualization of the OMC; (Y) - Greater than 50% of OMC visualized; N(Y) - Half of OMC visualized; (N) - Less than 50% of OMC visualized; N - No visualization of the OMC

| Subject ID | PRE LM | POST LM | PRE RSDI | POST RSDI | Failure/Success |
|---|---|---|---|---|---|
| 1 | 4 | 7 | 8 | 11 | F |
| 2 | 12 | 12 | 62 | 69 | F |
| 3 | 11 | 11 | 4 | 4 | F |
| 4 | 6 | 2 | 39 | 10 | S |
| 5 | 4 | 2 | 22 | 22 | S |
| 6 | 11 | 8 | 57 | 0 | S |

| Subject ID | Failure/ Success | Difference* in Total Volume | Difference in Main Nasal Cavity Volume | Difference in Sinus Volume |
|---|---|---|---|---|
| 1 | F | -5151.9 | -7580.7 | 2453.8 |
| 2 | F | -12871.9 | -2701.1 | -10178.1 |
| 3 | F | -5575.7 | -1848.6 | -4631.9 |
| 4 | S | 39104.1 | 5135.2 | 33853.3 |
| 5 | S | 13654.2 | 11104.2 | 1550.9 |
| 6 | S | 849.0 | -4822.1 | 5700.4 |

*All differences are (Post MMT – Pre MMT).

|  | Average Sinonasal Volume Difference* | Average Main Nasal Cavity Volume Difference | Average Paranasal Sinuses Volume Difference |
|---|---|---|---|
| **Failure Cohort** | -7866.5 ± 4339.9 | -4043.4 ± 3092.9 | -4118.7 ± 6331.5 |
| **Success Cohort** | 17869.1 ± 19472.7 | 3805.8 ± 8046.0 | 13701.5 ± 17574.8 |

*All differences are After MMT – Before MMT

| Failures | LHS | Score | RHS | Score | Total Score |
|---|---|---|---|---|---|
| 1 | (N) | 4 | N | 5 | 9 |
| 2 | (N) | 4 | N | 5 | 9 |
| 3 | Y | 1 | N | 5 | 6 |
| Successes | LHS | Score | RHS | Score | Total Score |
| 4 | N | 5 | (Y) | 2 | 7 |
| 5 | N(Y) | 3 | (Y) | 2 | 5 |
| 6 | Y | 1 | Y | 1 | 2 |

| Scoring System | |
|---|---|
| Y | 1 |
| (Y) | 2 |
| N(Y) | 3 |
| (N) | 4 |
| N | 5 |

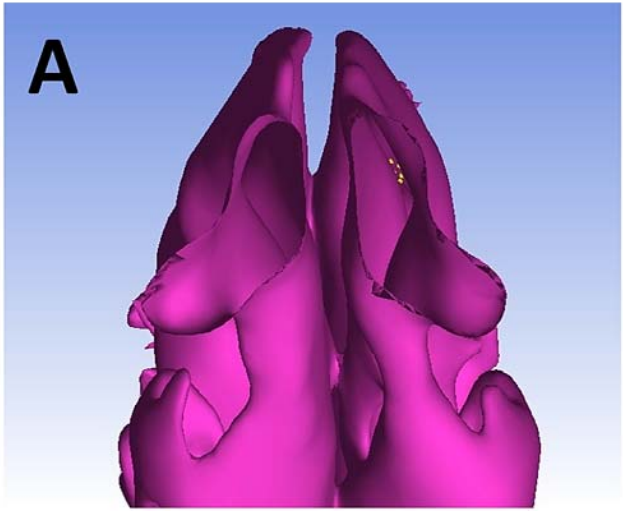
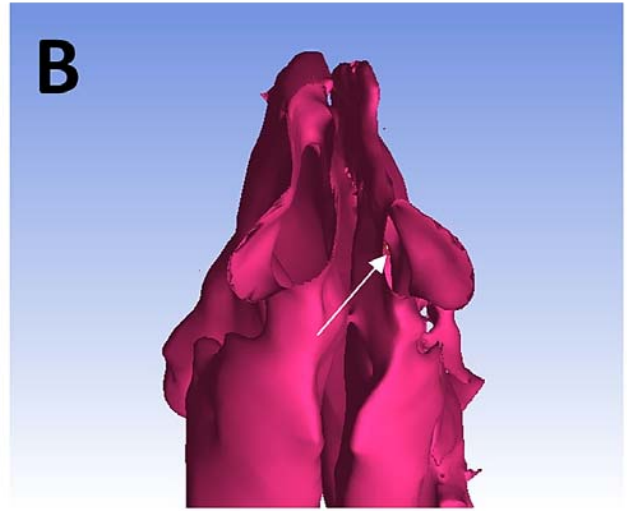
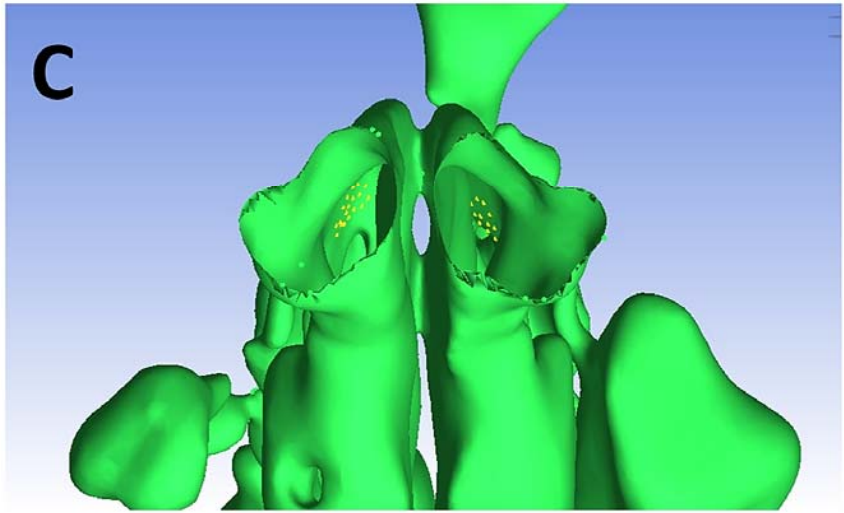

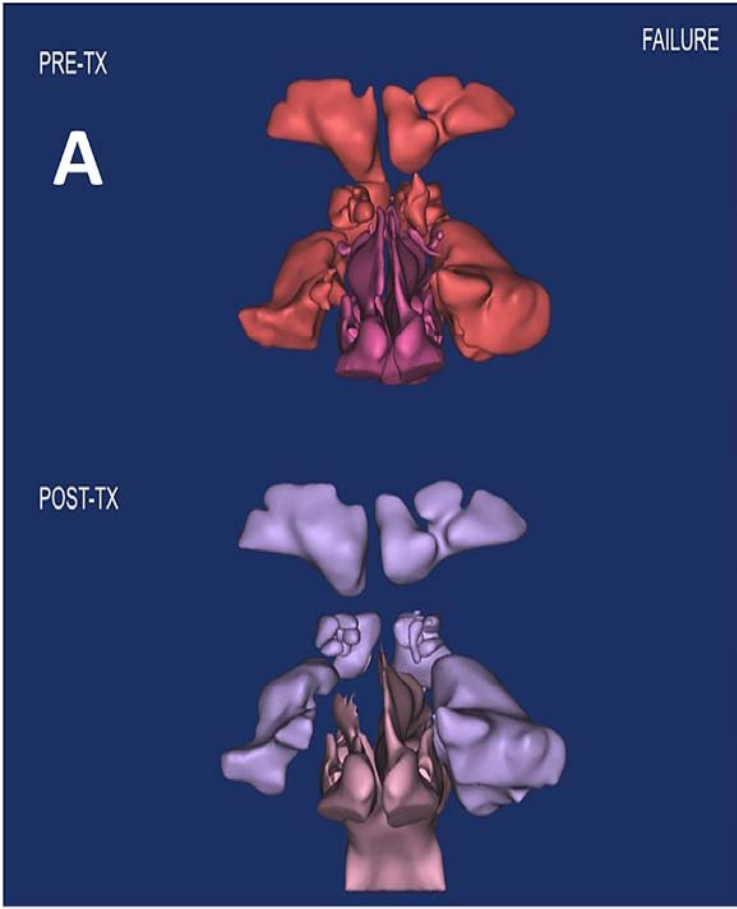 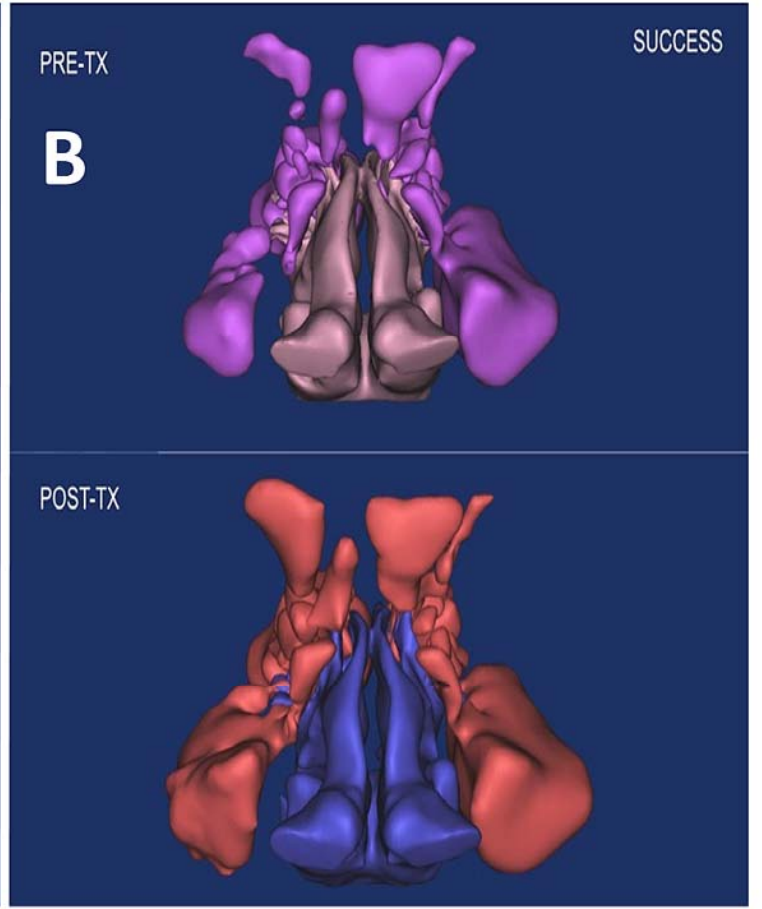

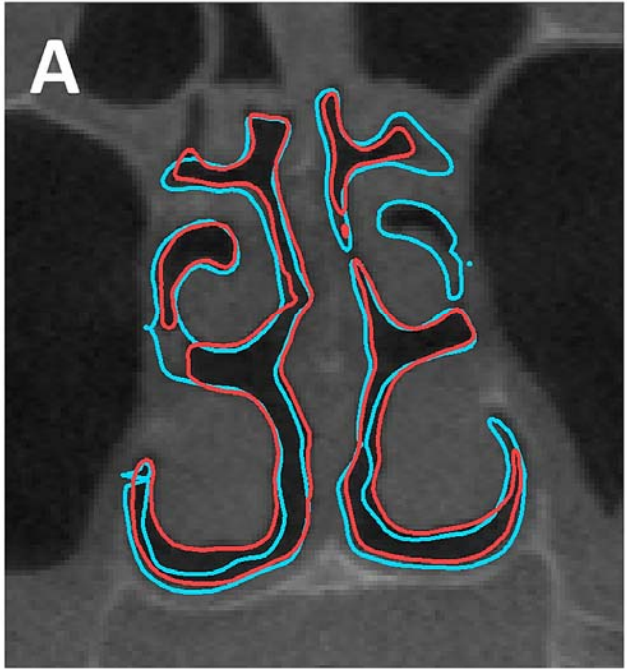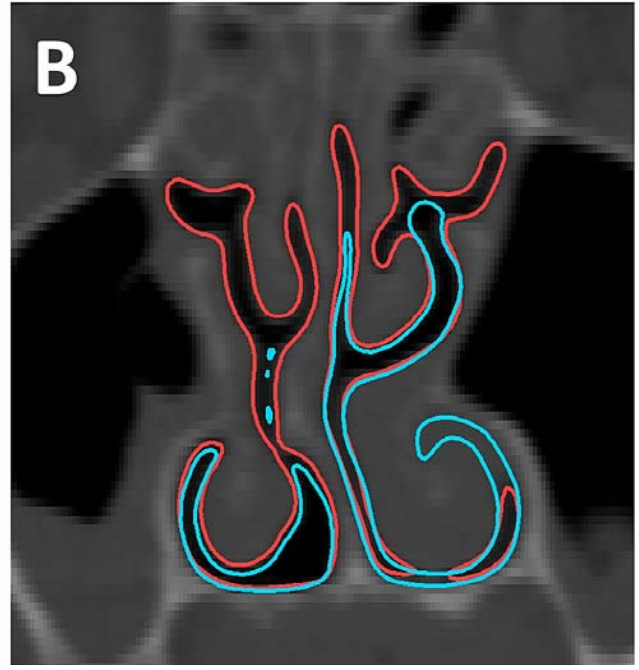

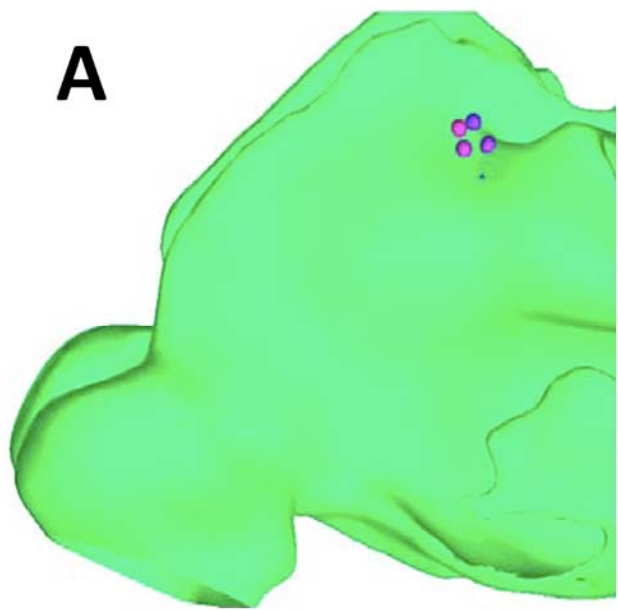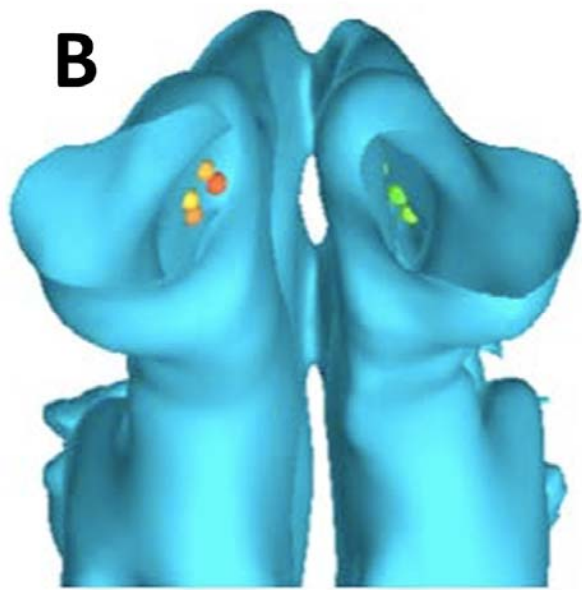